\documentclass[graybox]{svmult}

\usepackage{mathptmx}       
\usepackage{helvet}         
\usepackage{courier}        
\usepackage{type1cm}        
\usepackage{makeidx}         
\usepackage{graphicx}        
\usepackage{multicol}        
\usepackage{amssymb}   
\usepackage{amsmath}    
\usepackage[bottom]{footmisc}
\usepackage{cite}
\usepackage{todonotes}

\usepackage{changes}

\begin{document}

\title*{Regime change detection in irregularly sampled time series}
\author{Norbert Marwan, Deniz Eroglu, Ibrahim Ozken, Thomas Stemler, Karl-Heinz Wyrwoll, J\"urgen Kurths}
\institute{Norbert Marwan \at Potsdam Institute for Climate Impact Research (PIK), 14473 Potsdam, Germany, \\\email{marwan@pik-potsdam.de}
\and Deniz Eroglu \at Instituto de Ci\^encias Matem\'aticas e Computa\c{c}\~ao, Universidade de S\~ao Paulo, Brazil \\ Department of Mathematics, Imperial College London, United Kingdom \\  \email{deniz.eroglu@imperial.ac.uk}
\and Ibrahim Ozken \at Department of Physics, Ege University, 35100 Izmir, Turkey 
\and Thomas Stemmler\at School of Mathematics and Statistics, The University of Western Australia, Crawley, Western Australia 6009, Australia
\and Karl-Heinz Wyrwoll, \at School of Earth and Environment, The University of Western Australia, Crawley, Western Australia 6009, Australia
\and J\"urgen Kurths, \at Potsdam Institute for Climate Impact Research (PIK), 14473 Potsdam, Germany \\ Institute of Applied Physics of the Russian Academy of Sciences, 46 Ulyanova St., Nizhny Novgorod 603950, Russia.
}

\maketitle

\abstract{Irregular sampling is a common problem in palaeoclimate studies. We propose a method that provides
regularly sampled time series and at the same time a difference filtering of the data. The differences between 
successive time instances are derived by a transformation costs procedure. A subsequent recurrence analysis
is used to investigate regime transitions. This approach is applied on speleothem based palaeoclimate proxy 
data from the Indonesian-Australian monsoon region. We can clearly identify Heinrich events in the palaeoclimate
as characteristic changes in the dynamics.}

\section*{Introduction}
In the last decades, palaeoclimate research has experienced an exciting progress with ever-higher resolution 
and better age control high-resolution records, 
innovative technologies and types of proxies, as well as new data series analysis approaches, such as speleothem based proxies,
fluid inclusion analysis and laser ablation techniques, complex network based data analysis, etc.
\cite{Dennis2001,McDermott2001,Kennett2012,Rehfeld2013,McRobie2015}. This progress helps greatly to increase our understanding of
past climate variation and the mechanisms behind the climate system, but also to assess future climate-related
vulnerability of our society. Of particular interest are critical transitions, such as tipping points or regime shifts, because they 
can bring the climate system into another mode of operation \cite{Lenton2008,Scheffer2012}. Identifying tipping
points from measurements is no simple task. Several approaches have been proposed, such as testing for
slowing down and increase of the auto-correlation \cite{Scheffer2009}, reconstructing potentials of the dynamics by 
using the modality of the data distribution \cite{Livina2010}, using a modified detrended fluctuation analysis (DFA) \cite{Livina2007},
or the concept of stochastic resonance \cite{Braun2011}. While dynamical transitions are rather obvious when they 
appear in the first two moments (i.e., in mean or variance), they can be hidden when superimposed by signals of different 
time scales or by noise, issues frequently observed in palaeoclimate time series. For such problems, the application of methods 
from nonlinear time series analysis is a well accepted perspective, e.g., by using the fluctuation of similarity (FLUS) \cite{Malik2012}.
Another promising tool for the identification of subtle transitions is the framework of recurrence plots \cite{Marwan2007}.
Recurrence plots and their quantification consider the evolution of neighbouring states in a phase space. Besides characterizing
different classes of dynamics or testing for synchronization and nonlinear interrelationships and couplings of multiple systems, 
it allows to test for dynamical regime changes with respect to different properties, such as changes in the geometry of the
attractor, in the predictability of states, or in the intermittency behaviour \cite{Marwan2007,Donner2011,Eroglu2014}. The recurrence
plot framework has been successfully applied to investigate past transitions, e.g., in the Asian monsoon system \cite{Marwan2013} and 
in the East African climate \cite{Donges2011}, and to uncover a seesaw effect within the East Asian and Indonesian-Australian
summer monsoon system \cite{Eroglu2016}.

However, most palaeoclimate proxy records (independent of the actual archive) come with the challenge of
irregular sampling. While sampling in the field or in the lab is often done on a regular depth/length axis, varying 
sedimentation or growth rates result in variable time-depth relationships and in time series with non-equidistant
sampling points in the time-domain \cite{Breitenbach2012}. The most common procedure is data pre-processing using linear interpolation. However,
interpolation can lead to a positive bias in autocorrelation estimation (and, thus, an overestimation of the persistence time) 
and a negative bias in cross correlation analysis \cite{Rehfeld2011}. Therefore, several approaches have been suggested
for analyzing irregularly sampled time series \cite{Rehfeld2011,Scargle1982,Stoica2006,Rehfeld2014,Ozken2015}.

In the following we will focus on a recently proposed technique that is based on a measure that compares spike trains 
by quantifying the effort it needs to transform one spike train to the other one \cite{Victor1997,Hirata2009}. This measure corresponds to 
a modified difference filter (a common practice to remove low-frequency variation and trends), where we determine 
the differences by a criterion of how close subsequent short segments of an unevenly sampled time series are
by determining the cost needed to transform one segment into the following one \cite{Ozken2015}. 
Such comparison of successive segments has some similarity with the FLUS method \cite{Malik2012}, 
but instead uses the transformation cost as the similarity measure, and is thus directly applicable on irregularly sampled time series.
We illustrate this approach by analyzing a speleothem-based palaeoclimate record with respect to regime transitions.

\section*{Methods}
\subsection*{Transformation costs time series (TACTS)}

Cumulative trends or low-frequency variations are common in palaeoclimate proxy records, but are often undesirable and can cause
difficulties in the analysis. One frequently used solution is the difference filter, where the values of the 
proxy record are replaced by the differences of subsequent values, $y(t-\Delta t /2)=x(t)-x(t-\Delta t)$, with $\Delta t$ the
sampling time of a regularly sampled time series.
Another, even more challenging problem is the irregular sampling frequently occurring in palaeoclimate proxy records.
The {\it transformation costs time series} (TACTS) approach tries to overcome both problems by transforming irregularly sampled time series to regular 
ones and simultaneously using the transformation cost as the difference value. This procedure induces less loss of information 
compared to traditional interpolation procedures.

The core of the TACTS method is to measure the shortest distance (transformation cost) between two data segments by 
using two different processes: (i) {\it shifting points} in time which causes changes in the amplitude for marked data and 
(ii) {\it adding-deleting} operations. The process starts with dividing the data into small and equally sized segments.  
These segments can have different number of points, because the points are not equally sampled. The transformation 
costs between all sequence windows are then calculated by
\begin{align}
p(S_a,S_b) = \overbrace{\sum_{(\alpha,\beta)\in{C}} \lbrace \lambda_0|t_a(\alpha)-t_b(\beta)| +  \lambda_k|L_{a}(\alpha)-L_{b}(\beta)|\rbrace}^\text{shifting} 
+\underbrace{\lambda_S(|I|+|J|-2|C|)}_\text{adding/deleting}.
\label{eq:cost}
\end{align}

The equation states two distinct operations for two essential processes.
If the operation is {\it shifting} then the first part of the equation involves, otherwise the {\it adding-deleting} operation involves as 
the second part. In the first part, the summation is over the pairs $(\alpha,\beta)\in C$, where $C$ is the set of points that will be 
shifted in time and changed in amplitude. $\alpha$ and $\beta$ are the $\alpha$th event in the first segment ($S_a$) and the $\beta$th 
event in the second segment ($S_b$). The amplitude of points which are $\alpha$th and $\beta$th elements of $S_a$ and $S_b$ 
are denoted by $L_{a}(\alpha)$ and  $L_{b}(\beta)$ respectively. The data-adapted constants $\lambda_0$ and $\lambda_k$ are given by
\begin{subequations}
\begin{align}
\label{eq:lambda0}
\lambda_0&= \frac{M}{\text{total time}} \\
\label{eq:lambdak}
\lambda_k&= \frac{M-1}{\sum_i^{M-1}|x_i-x_{i+1}|}.
\end{align}
\label{eq:lambda}
\end{subequations}
where $M$ is the total number of events, and $x_i$ is the amplitude of $i$th element in the time series. 

In the second part of Eq.~(\ref{eq:cost}),  $I$ and $J$ are sets of indices of the events in $S_a$ and $S_b$, respectively. 
The parameter $\lambda_S$ is the cost of deleting or adding processes and is used as an optimization parameter. The 
selection of optimum $\lambda_S$ is the following: first we calculate total cost time series for the entire range of $\lambda_S\in[0,4]$ 
with step size $\Delta \lambda_S = 0.01$. Then we examine frequency distributions for each cost time series. 
Since each cost value is independent of the others, we expect to have a normal distributed histogram and 
choose the optimal $\lambda_S$ according to the best fit on normal distribution. 

Eq.~(\ref{eq:cost}) is a metric distance function, satisfying the following three conditions:  
\begin{itemize}
\item $p(S_a,S_b)\ge0$ (positive)
\item $p(S_a,S_b)=p(S_b,S_a)$ (symmetric)
\item $p(S_a,S_c)\le p(S_a,S_b)+p(S_b,S_c)$ (triangle inequality)
\end{itemize}

\begin{figure}[h]
\centerline{\includegraphics[width=0.8\linewidth]{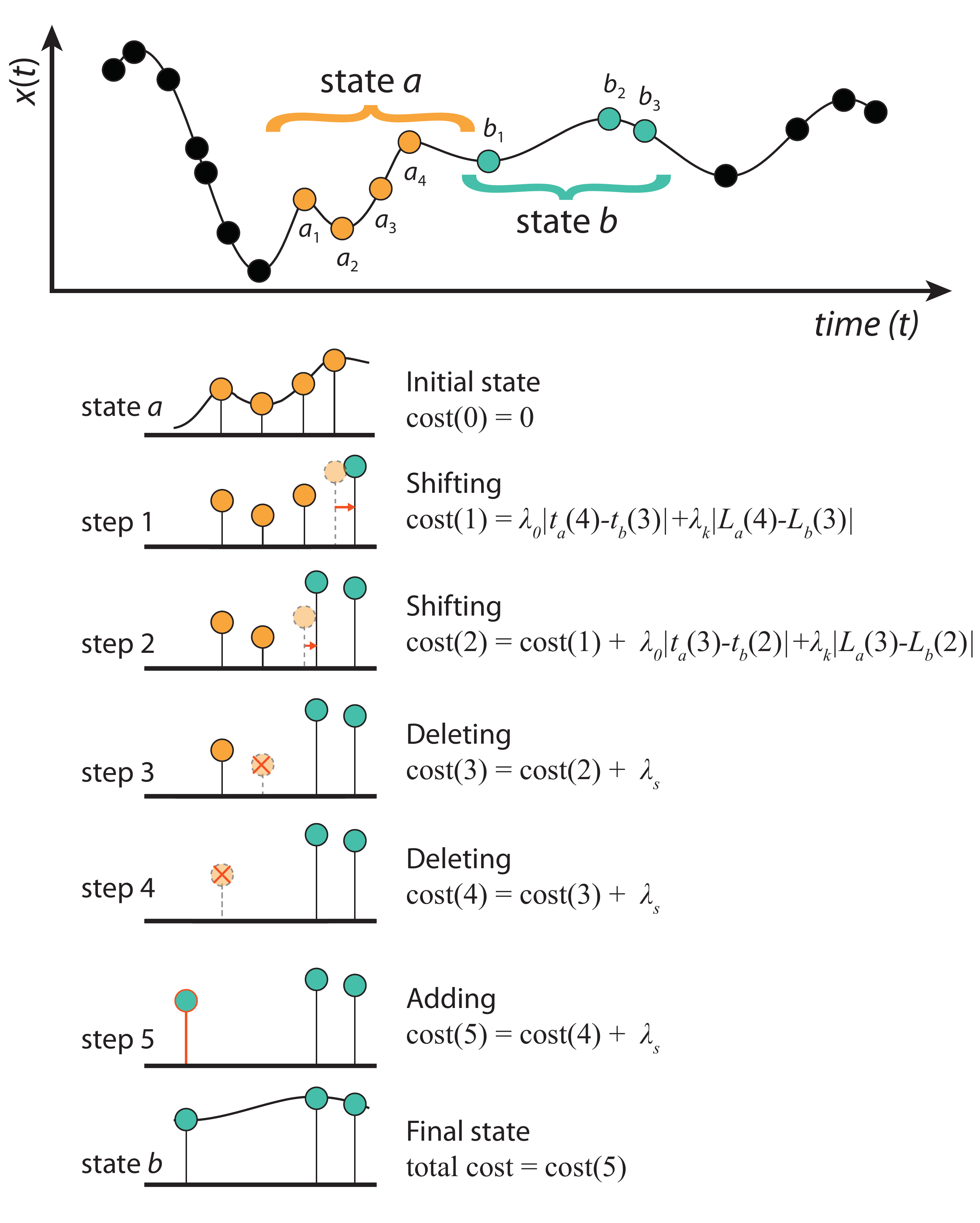}}
\caption{Illustration of the transformation cost time series method, which finds the minimum transformation cost between two data 
segments such as state $a$ and state $b$ in the top panel. In five steps state $a$ is transformed into state $b$. At steps $1$ and $2$, 
we apply {\it shifting} a point in time and, as a consequence of shifting, changing the amplitude of the point. These operations cost 
regarding to first part of Eq.~(\ref{eq:cost}). Steps 3 and 4 are {\it deleting} and step 5 is {\it adding} a point; each of these operations costs 
a constant $\lambda_s$. The costs are written next to the related processes according to Eq.~(\ref{eq:cost}).}
\label{fig:tacts}
\end{figure}

Now we illustrate the method for two consecutive segments. Irregularly sampled data is equally spaced into small windows 
which are given as state $a$ ($S_a = \{a_\alpha\}_{\alpha=1}^4$) and state $b$ ($S_b = \{b_\beta\}_{\beta=1}^3$). 
The costs computed between the states and all details are given in Fig.~\ref{fig:tacts} step by step.

Note that the decision of which operation process to minimizes costs is important.
The transformation by shifting costs $\lambda_0|t_a(\alpha)-t_b(\beta)|+\lambda_1|L_{a}(1)-L_{b}(1)|$ and deleting 
and adding a point costs $2\lambda_S$. Here we chose the least cost operation to either shift or delete/add. 
Therefore, in the algorithm, we consider all these possibilities and chose the operation carefully.

The final appearance of the cost time series is as follows: assume that we have an irregularly sampled time series  
$\{u_i\}_{i=1}^N$,  where $N$ is the number of points. The data is divided to a set of $W$-sized $n$ segments and each 
segment has a minimum of a certain number of points, therefore, 
$$
TACTS = \{p(W_i,W_i+1)\}_{i=1}^{n-1}
$$
for all sequence windows. This leads to an equally sampled and detrended time series. 
The resulting cost values series can be considered as the difference filtered time series with a regularly sampled time axis
and can be further analysed with standard or advanced time series analysis tools, e.g., in order detect
regime shifts (Fig.~\ref{fig:tacts}).

\subsection*{Recurrence analysis}
\begin{figure}[h]
\centering{\includegraphics[width=0.5\linewidth]{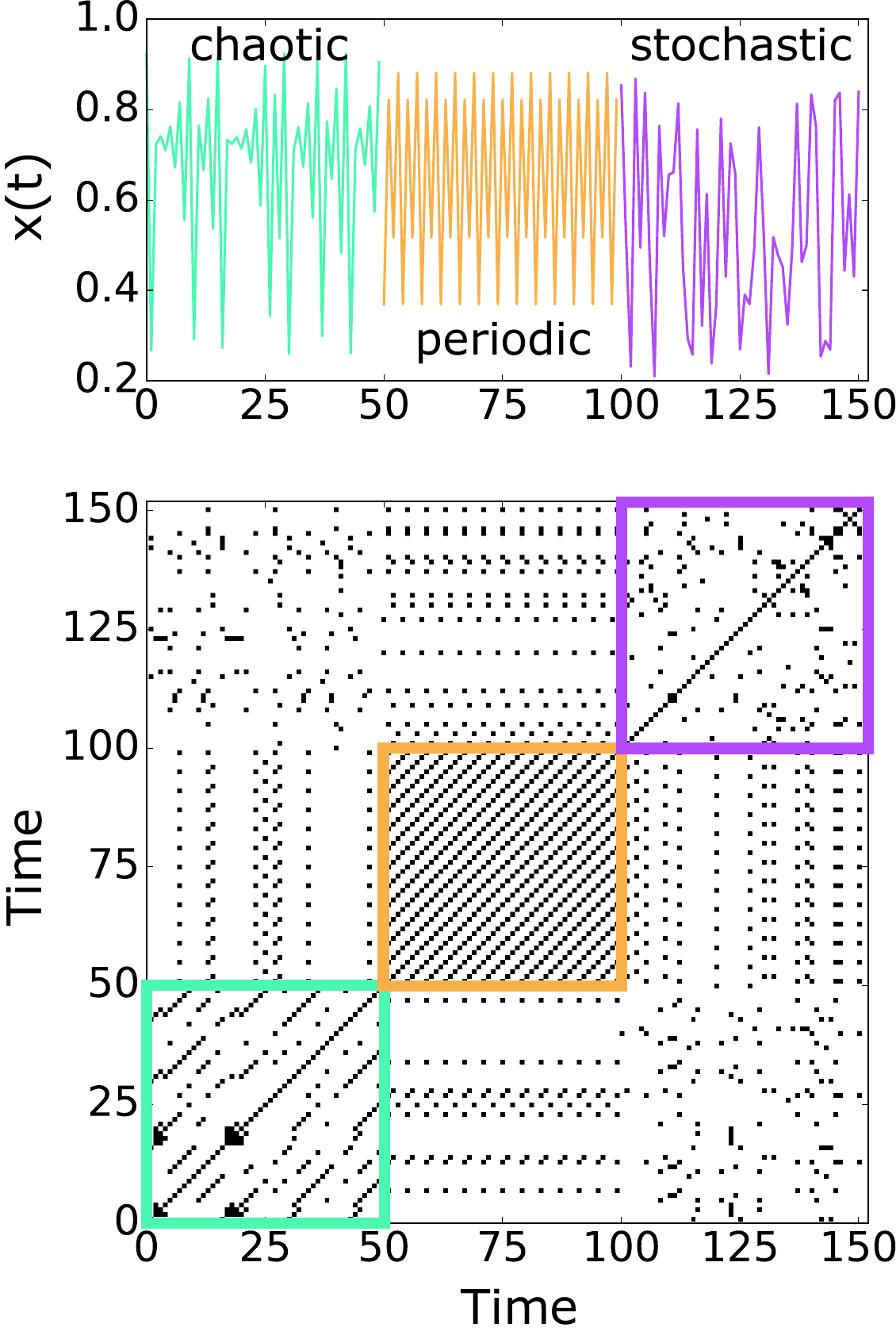}}
\caption{Example of a recurrence plot for changing dynamics from chaotic via periodic to stochastic dynamics,
each lasting $50$ time steps. In the periodic region, continuous long diagonal lines are observed, in the
chaotic region, shorter diagonals and single points appear, and in the stochastic part, we find almost only single points. }
\label{fig:rp}
\end{figure}

Recurrence is an ubiquitous property of many dynamical systems. Slight changes in observed recurrence behaviour allow to 
infer changes in the dynamics \cite{Marwan2007,Marwan2011}. In order to investigate recurrence properties, recurrence
plots and recurrence quantification analysis have been developed \cite{Marwan2007,Marwan2008}. A recurrence plot is the graphical representation of
those times $j$ at which a system recurs to a previous state $\vec{x}_i$:
\begin{equation}
R_{i,j} = \Theta(\varepsilon - \| \vec{x}_{i} -  \vec{x}_{j} \| ), \qquad i,j = 1,\ldots, N
\end{equation}
with $\Theta$ the Heaviside function, $\varepsilon$ a recurrence threshold, $\| \vec{x}_{i} -  \vec{x}_{j} \|$ the Euclidean
distance between two states $\vec{x}_i$ and $\vec{x}_j$ in the phase space, and $N$ the number of observations (or time
series length). Such a recurrence plot consists of typical large-scale and small-scale features that can be used to interpret the
dynamics visually. Important features are diagonal lines: similar evolving epochs of the phase space trajectory cause
diagonal structures parallel to the main diagonal in the recurrence plot. The length $l$ of such diagonal line structures depends 
on the dynamics of the system (periodic, chaotic, stochastic) Fig.~\ref{fig:rp} and can be directly related with dynamically invariant 
properties, like $K_2$ entropy \cite{Marwan2007}. Therefore, recurrence quantification analysis (RQA) uses the features 
within the recurrence plots for defining measures of complexity. For example, the distribution $P(l)$ of line lengths $l$ is used by several
measures of complexity in order to characterise the system's dynamics in terms of predictability/determinism or laminarity.
The measure {\it determinism} DET is the fraction of recurrence points (i.e., $R_{i,j}=1$) that form diagonal lines and
can be computed by 
\begin{equation}\label{eq:det}
DET = \frac{\sum_{l_{\min}}^N l \cdot P(l)}{\sum_{i,j=1}^N R_{i,j}}.
\end{equation}

In order to study the time-dependent behaviour of a system or time series,  RQA measures can be computed within a moving window, 
applied on the time series. The window has size $w$ and is moved with a step size $s$ over the data in such a way 
that succeeding windows overlap with $w - s$. This technique can detect chaos-period 
and also more subtle chaos-chaos transitions \cite{Marwan2007}, or different kinds of transitions 
between strange non-chaotic behaviour and period or chaos \cite{Ngamga2007}. 
Moreover, the reliability of several RQA measures was investigated by their scaling properties with respect 
to critical points in the dynamics \cite{Afsar2015}.

\section*{Palaeoclimate regime transition}

\begin{figure}[b!!!]
\centerline{\includegraphics[width=.9\linewidth]{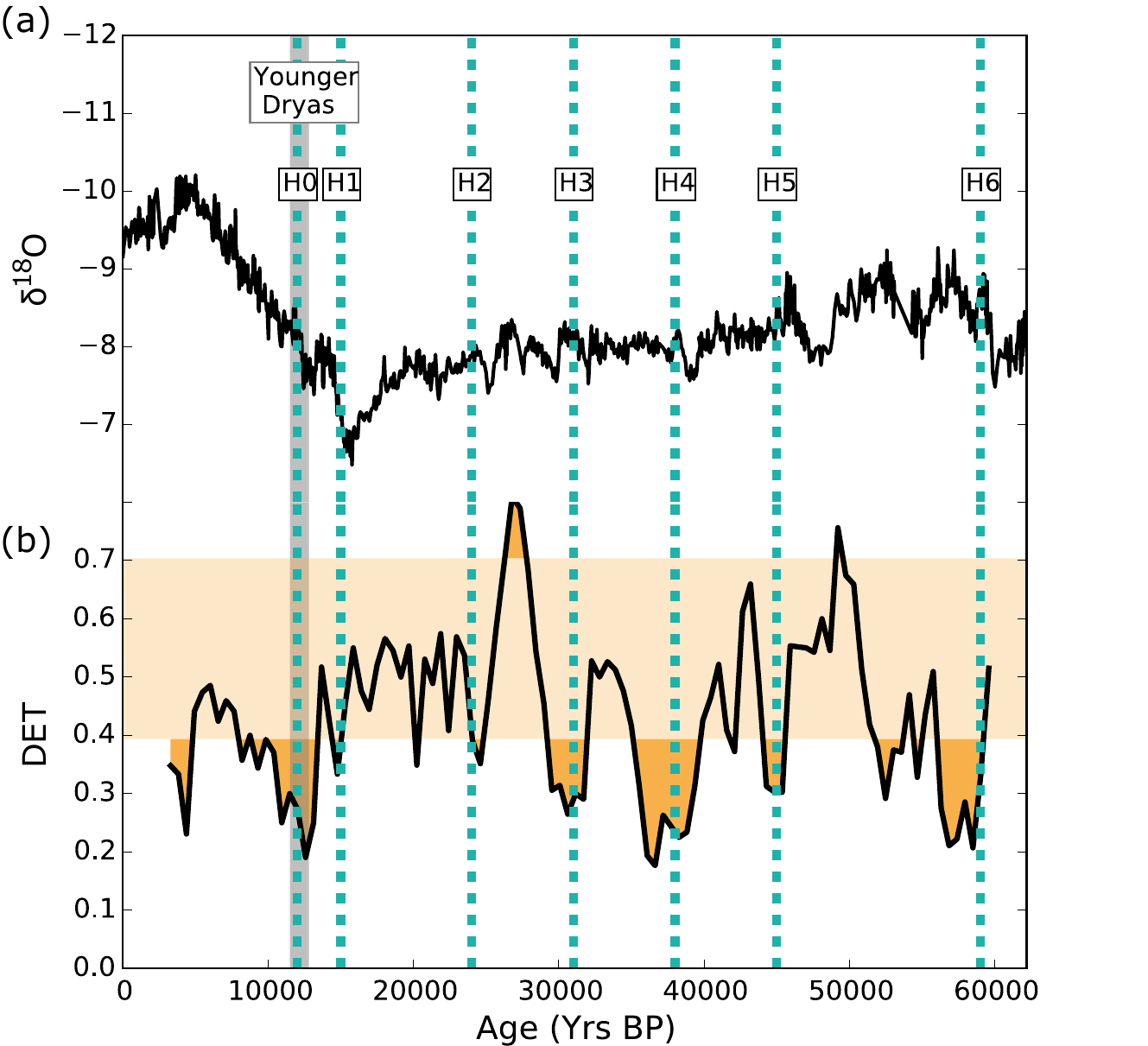}}
\caption{(a) $\delta^{18}$O record of Secrete Cave, Borneo. (b) RQA-determinism $DET$, Eq.~(\ref{eq:det}), time series resulting from the transformation cost time series. The light orange band of the $DET$ indicates the 90\% confidence interval. The vertical lines H1--H6 give the six Heinrich events as well as H0, the Younger-Dryas. }
\label{fig:results}
\end{figure}

To illustrate the power of the techniques we advocate here, we choose as illustrating example a speleothem 
$\delta$$^{18}$O record from the Secret Cave at Gunung Mulu in Borneo/ Indonesia~\cite{Carolin2013}. 
This particular record has been interpreted as a time series of the dynamics of the East Asian--Indonesian 
-- northwest Australia monsoon. This monsoon regime provides a circulation regime
that strongly links both hemispheres and serves as a major heat source, playing a significant 
role at planetary scale \cite{mcbride1987,chang2006}. Central to its geography is the Maritime Continent 
which provides a core region of monsoon activity \cite{Ramage1968,chang2004}. A transect in regional 
precipitation patterns from the northern part of the Maritime Continent to the northern margin of Australia 
coincides with a change from the dominance of the boreal summer monsoon to the austral summer monsoon
\cite{chang2006,chang2004,Robertson2011}. The transect captures key palaeoproxy monsoon records and 
has the potential to provide details of the function of the monsoon regime over Quaternary time scales 
\cite{ayliffe2013,Carolin2013,denniston2013,Partin2007}. Imbedded in some of these records are short-lived 
millennial and centennial scale events, and, more general, relatively short-lived phases of climate instability. 

While the full proxy record is around 100,000 years, we only analyze the last 62,000 years of the $\delta$$^{18}$O 
record (Fig.~\ref{fig:results}(a)). Before the 62,000 years many gaps appear and the data become too sparse to give any useful information 
about. The record used for the analysis contains about $1,200$ data points. Time intervals between measurements are irregular 
and follow a Gamma distribution with a skewness of 4.9. In our analysis we use a window length of $\approx 210$ 
years to calculate the TACTS. While the parameters $\lambda_{0,k}$ are determined by 
Eq.~(\ref{eq:lambda}), we optimize $\lambda_S=1.07$. 

The next step is to analyze the regularly sampled 
TACTS with RQA using a sliding window method. We consider 30 data points (or 6,200 years) of the 
TACTS as our window size. Given the average number of points in the proxy record, 
30 data points of the TACTS correspond to approximately 100 to 140 points in the 
original proxy. Using an overlap of 90\% of consecutive windows, we determine the $DET$ (Eq.~(\ref{eq:det})) 
for each window with length of 6,200 years (Fig.~\ref{fig:results}(b)). 
The recurrence threshold is selected to be $\epsilon=20\%$ of the standard deviation of the data in the particular window. The advantage 
of this $\epsilon$ selection scheme is that it allows us to analyze proxy records with inherent non--stationarity. 
In addition, we determine the 
statistical significance of $DET$ using the bootstrapping method as outlined in~\cite{Marwan2013}
(light red band in Fig.~\ref{fig:results}(b)). 

The determinism $DET$ indicates several distinct regime changes in the time series from less to more
predictable (and vice versa) dynamics (Fig.~\ref{fig:results}(b)).
Most minima of $DET$, signified as periods of decreased predictability,
coincide with the so called Heinrich events (H1 to H6). Heinrich 
events are identified in the North Atlantic sediments as layers of ice-rafted debris, associated with the 
coldest phase just before the Dansgaard-Oeschger Events, and result from episodic discharge of icebergs in the Hudson Bay 
region \cite{Clement2008,McNeall2011}.

Heinrich events are well represented in the Chinese speleothem and loess record
as periods of weakened summer monsoon and intensified winter monsoon. \cite{An2014}.  In their interactions 
with the Siberian Mongolian High of the East Asian Winter Monsoon they can be expected to trigger
cold surges which leave their imprint in the proxy palaeoclimate record \cite{Wyrwoll2016}.  During the 
East-Asian Winter Monsoon (EAWM), the 
Siberian High with its central pressure reaching in excess of 1035~hPa, dominates much of the Eurasian continent.  
Strong northwesterly flows occur at its eastern margins, where one branch of the flow separates and first is directed 
eastward into the subtropical western Pacific and then tends southward in the direction of the South China Sea.
These cold 
air `excursions', also described as `cold surges', are channeled by the trough southwards and are a characteristic 
feature of the EAWM \cite{lau1987}. Their path is in part related to relief controls of the Tibetan Plateau. Cold 
surges transport absolute vorticity and water vapour up-stream of the South China Sea  to the Equator 
\cite{Koseki2013} and lead to the flare-up of convective activity over the Maritime Continent \cite{chan2004}. 
In the Borneo region, cold surges enhance surface cyclonic circulation triggering  the Borneo Vortex, 
which leads to deep convection giving rise to heavy rainfall events \cite{Koseki2013,Ooi2011}.

It is noteworthy that in raw $\delta$$^{18}$O record from the Secret Cave the Heinrich events are almost
indistinguishable from other variations in the time series. 
In the original work by Carolin \emph{et al.}, H1 to H6 were detected by visual comparison of the
record to others (e.g., NGRIP), 
but the Younger Dryas (coinciding with the H0 event), was not detected \cite{Carolin2013}. However, our method 
clearly extracts these events, including the previously not detected Younger Dryas, and highlights the hidden impact of such distal forcing. Moreover,
it allows an objective, quantitative analysis, while Carolin \emph{et al.} rely on the subjective method of matching 
extreme proxy values with specific dates. At present, the Borneo Vortex leaves a strong climate signal on the regional 
precipitation patterns \cite{Ooi2011}. We propose that the prominence of the `instability climate 
phases', coincident with the timing of Heinrich events in the Borneo record, is an expression of regional controls that are linked 
to the operation of the Borneo Vortex. 
The claim draws attention to the need to give more consideration to specific 
regional controls in explaining  the palaeoclimate proxy record rather than simply appeal to global or hemispheric 
controls.

\section*{Conclusion}

We have used the Secret Cave $\delta^{18}$O record from Borneo to illustrate the usefulness of the novel TACTS method
for analyzing palaeoclimate records. 
TACTS can transform irregularly sampled time series into a regularly sampled cost time series. This is an important step, 
since most modern time series analysis methods -- like the RQA used here -- require a regular sampled time series as an 
input. Furthermore, the TACTS method is less biased than interpolation methods frequently used to 
transform irregularly sampled into regularly sampled data sets. This transformation only requires three parameters. 
The two parameters $\lambda_{0,k}$ are given by the average amplitude and frequency of the record (see Eq.~(\ref{eq:lambda})), 
while $\lambda_S$ needs to be optimized. Being a difference filter, the TACTS method lends itself naturally for 
palaeoclimate investigations, where proxy records often have some non-stationarity and usually need to be 
detrended. As we have shown the detrending is build into the TACTS method, therefore we do not need this 
additional step in our time series analysis.

Applying the TACTS and RQA approach on palaeoclimate data from the Secret Cave speleothem,
we were able to identify regime changes 
in the monsoon activity during the last 62,000 years. We report on several distinct regime changes coinciding with 
the Heinrich events H1 to H6 and therefore add quantitative evidence of these impacts to previous, more qualitative 
studies \cite{Carolin2013}. Moreover, our analysis clearly unveils that also the Younger Dryas had an impact on 
the monsoon activity over the Maritime Continent.   

Given that irregular sampling of proxy records is quite common in Earth science, the TACTS method has 
large potential in quantitative Earth science without prior modification or preprocessing the data. 

\begin{acknowledgement}
This work was supported by grants from the Leibniz Association, grant SAW-2013- IZW-2 (Gradual environmental change versus single catastrophe -- Identifying drivers of mammalian evolution) and the European Union's Horizon 2020 Research and Innovation programme under the Marie Sk{\l}odowska-Curie grant agreement No 691037 (RISE project QUantitative palaeoEnvironments from SpeleoThems QUEST).
We thank Sebastian Breitenbach for fruitful discussions and support.
\end{acknowledgement}

\end{document}